\documentclass[sigconf,screen]{acmart}

\usepackage{fancyvrb}
\usepackage{graphicx}
\usepackage{relsize}
\usepackage{enumitem}
\usepackage{booktabs}
\usepackage{url}
\usepackage{array,multirow,tabularx,ragged2e}
\usepackage[english]{babel}
\usepackage{verbatimbox}
\usepackage{amsmath,amsfonts}
\usepackage{algorithmic}
\usepackage{graphicx}
\usepackage{textcomp}
\usepackage{xcolor}
\usepackage{pdflscape}
\usepackage{microtype}
\usepackage{multirow}

\usepackage{siunitx}
\newcolumntype{C}{>{\Centering\arraybackslash}X}
\def\BibTeX{{\rm B\kern-.05em{\sc i\kern-.025em b}\kern-.08em
    T\kern-.1667em\lower.7ex\hbox{E}\kern-.125emX}}
\AtBeginDocument{%
  \providecommand\BibTeX{{%
    \normalfont B\kern-0.5em{\scshape i\kern-0.25em b}\kern-0.8em\TeX}}}

\setcopyright{acmlicensed}
\acmPrice{15.00}
\acmDOI{10.1145/3558489.3559068}
\acmYear{2022}
\copyrightyear{2022}
\acmSubmissionID{fsews22promisemain-p5-p}
\acmISBN{978-1-4503-9860-2/22/11}
\acmConference[PROMISE '22]{Proceedings of the 18th International Conference on Predictive Models and Data Analytics in Software Engineering}{November 17, 2022}{Singapore, Singapore}
\acmBooktitle{Proceedings of the 18th International Conference on Predictive Models and Data Analytics in Software Engineering (PROMISE '22), November 17, 2022, Singapore, Singapore}

\begin{document}

\title[Feature Sets in Just-in-Time Defect Prediction: An Empirical Evaluation]{Feature Sets in Just-in-Time Defect Prediction:\\An Empirical Evaluation}

\author{Peter Bludau}
\email{bludau@fortiss.org}
\orcid{0000-0003-4738-0488}
\affiliation{%
  \institution{fortiss - Research Institute of the Free State of Bavaria}
  \city{Munich}
  \country{Germany}
}

\author{Alexander Pretschner}
\email{alexander.pretschner@tum.de}
\orcid{0000-0002-5573-1201}
\affiliation{%
  \institution{Technical University of Munich}
  \city{Munich}
  \country{Germany}
}

\begin{abstract}
Just-in-time defect prediction assigns a defect risk to each new change to a software repository in order to prioritize review and testing efforts. Over the last decades different approaches were proposed in literature to craft more accurate prediction models. However, defect prediction is still not widely used in industry, due to predictions with varying performance. 
In this study, we evaluate existing features on six open-source projects and propose two new features sets, not yet discussed in literature. 
By combining all feature sets, we improve MCC by on average 21\%, leading to the best performing models when compared to state-of-the-art approaches. 
We also evaluate effort-awareness and find that on average 14\% more defects can be identified, inspecting 20\% of changed lines.
\end{abstract}



\maketitle


\section{Introduction}

Finding defects early in the software development process is important as it reduces effort and costs for software development teams.
Many software engineering best practices are developed around the idea to assist developers to find and fix defects in a timely manner.
Defect prediction performs an automated identification of defects, traditionally for every module or release of a system.
Just-in-time (JIT) defect prediction instead classifies whether new changes are error-prone.
Therefore, it helps the developer to focus reviewing and testing efforts to particularly risky commits and to focus on smaller fractions of the software system, when reviewing changes. 
In order to learn the relations between clean and bug introducing changes machine-learning
is widely used in defect prediction studies. 
When applying JIT defect prediction in practice, researchers found that the effort spent by a developer to identify a defect
is especially important as it helps to prioritize more defect-prone commits.
Effort-aware approaches focus on minimizing inspected lines of code while finding most defects.
These models are more practitioner-oriented, as they take into account a limited testing contingent.
Researchers have proposed specialized unsupervised \cite{Yang2016} and supervised \cite{Kamei2013a, Huang2019c} approaches to improve effort-awareness.

In recent years, deep learning approaches are more often utilized to find more syntactic and semantic relations and hence to improve defect prediction models.
However, studies found that metric-based approaches may outperform such approaches in predictive software engineering tasks such as defect prediction \cite{Fu2017a, Pornprasit2021ARXIV, Majumder2018}. 
Therefore, researchers call for investigating
novel more effective features to further enhance metric-based JIT defect prediction \cite{Zeng2021, Pornprasit2021ARXIV}. 

Different projects apply deviant development and change integration techniques, that lead to varying project characteristics determining the defect proneness of changes. 
Therefore, it is not easy to identify the most expressive features for a specific project.
However,selecting only the important features for JIT defect prediction 
mitigates over-fitting and allows more effective review support. 

In this study, we investigate the effectiveness of state-of-the-art JIT defect prediction features in terms of predictive performance and effort-awareness. 
We follow the call from various researchers and propose two new sets of features from additional sources, 
which were not yet discussed in literature.
These novel features incorporate attributes from the software development workflow and more fine-grained information based on abstract syntax tree (AST) changes.
We perform an thorough empirical evaluation of state-of-the-art features and the novel feature sets on six open-source projects and compare their performance in the context of defect prediction.
We further investigate which features lead to a higher predictive performance, save effort
and are most important given different development environments. 

Our contributions can be summarized as follows:
\begin{itemize}[leftmargin=*]
    \item We propose two new features sets for JIT defect predictions. 
    One is based on metrics from the software development workflow. The other utilizes change information extracted from the AST.
    \item The novel features improve JIT defect prediction performance models on all investigated projects in terms of standard and effort-aware evaluation metrics in a cross- and a time-dependent validation setting. 
    Features from the AST change (ASTA, DEPTHA) are under the top-3 most important features in 5 of 6 projects.
    \item We publish our model and evaluation code as well as more detailed results with respect to six open-source projects.
\end{itemize}

\section{Background and Related Work} \label{RelatedWorkRefs}

This study investigates the effectiveness of JIT defect prediction models based on state-of-the-art approaches and novel feature sets. 
In the following, we introduce the background and related work.

\subsection{Just-in-Time Defect Prediction}

Traditional defect prediction models focus on identifying defective modules or files on a release basis and therefore whole modules or class files are predicted as defect-prone.
In order to offer more direct feedback, Mockus and Weiss \cite{Mockus2000} propose defect prediction on a change basis, predicting defect-inducing changes whenever a new change is submitted.
In recent years, such JIT defect prediction is more actively studied, as it provides several benefits over traditional defect prediction.
First, the prediction is available early after the initial development task is accomplished. 
This leads to faster reviewing activities, as the changes are still fresh on the developer's mind \cite{Shihab2012}.
Second, with predictions on change-level, reviewing efforts are reduced to the changed lines within the changed files,
leading to smaller and more targeted reviews \cite{Kamei2013a}.

Many studies evaluated the effectiveness of JIT defect prediction on open-source and industrial projects.
Śliwerski et al. \cite{Jaceksliwerski2005} studied defect-inducing changes in two open-source projects. Their approach led to the conclusion that changes committed on Fridays are more likely to introduce defects.
In a similar observation, Eyolfson et al. \cite{Eyolfson2011} found that the time of day and experience of the developer has an influence on the defect-proneness of changes.
Several studies \cite{Aversano2007, Kim2008, Kamei2013a} extended the initial set of code change properties from Mockus and Weiss \cite{Mockus2000} utilizing change-log information and change metadata. They report that the addition of features extracted from the version control system and the issue tracking system can elevate the predictive performance of JIT defect prediction models.
Later, Kononenko et al. \cite{Kononenko2015} investigated the role of code reviews in JIT software defect prediction. 
They extracted properties from code review systems and used them as additional features in their experiments, contributing significantly to the explanatory power.
A combination of the process and review features shown above, are often used in recent JIT defect prediction studies (e.g., \cite{Hoang2019, Nayrolles2018, Huang2017}).
Similarly, we use the aforementioned features as the state-of-the-art baseline feature set described in Section \ref{ApproachSOTA} and shown in Table \ref{tab:featuresSotA}.
Our study sets out to investigate additional features from software development practices and evaluate these models on a recently published defect data set \cite{bludau2022}. 

Recently more advanced modeling techniques, such as ensemble learning (e.g., \cite{Yang2017, Young2018}) and deep learning (e.g., \cite{Yadav2018, Hoang2019}) are proposed in the context of JIT defect prediction. 
We do not study such techniques in this work. 
However, we compare the JIT defect prediction results obtained in this study with deep learning approaches in order to validate the findings from recent studies \cite{Zeng2021, Pornprasit2021ARXIV}, stating that deep learning might not be able to outperform simpler metric-based approaches in regards to learning costs and predictive performance.

\subsection{Effort-Awareness in JIT Defect Prediction}

For developers with a fixed time-budget it is especially important to use cost-effective techniques when performing verification and validation activities \cite{Arisholm2010}.
Kamei et al. \cite{Kamei2013a} were the first to focus on the effort-awareness of JIT defect prediction models. 
Inspired by Ostrand et al. \cite{Ostrand2005}, finding that the majority of defects is contained in about 20\% of code, they propose to evaluate effort-aware models by the proportion of defects they can find with only 20\% of lines inspected (effort).
Therefore, in an effort-aware scenarios all commits are ranked by their probability to introduce a defect and the higher ranked commits are inspected first.
They design an effort-aware linear regression model (EALR) and evaluate it on six open source and five commercial projects. 
They show that effort-aware JIT defect prediction can reveal defect-prone changes early and thus reduces the effort required for software tests and code reviews.

Many studies emerged analyzing supervised and unsupervised effort-aware JIT defect prediction approaches \cite{Fu2017, Huang2017, Huang2019c, Yang2016}.
Based on latest results, \textit{CBS+} proposed by Huang et al. \cite{Huang2019c} is the best performing approach, which outperforms \textit{EALR} \cite{Kamei2013a} and unsupervised approaches like \textit{LT} \cite{Yang2016} and \textit{CHURN} \cite{Liu2017}.
\textit{CBS+} combines a supervised model \cite{Kamei2013a} and an unsupervised approach \cite{Yang2016} by first classifying all commits in two groups (defect-prone or not) and then ranking them based on their potential defect-proneness and size. 
Since, smaller commits are less error-prone \cite{Moser2008a}, \textit{CBS+} uses the ratio between the predicted defect-proneness and the size of the change and sorts them in descending order. 
Yan et al. \cite{Yan2020} confirmed that \textit{CBS+} is the best performing effort-aware approach also in an industrial context.
However, the most important model features are not stable when comparing their industrial projects to open-source projects, leading to the conclusion that development workflow and project characteristics do also determine feature importance.
 
Inspired by the results shown above we investigate the potential of novel features in an effort-aware setting. We perform additional evaluations to compare supervised and unsupervised JIT defect prediction models with different feature sets.

\begin{table}
 \centering\small
 \caption{State-of-the-art features for JIT defect prediction}
 \label{tab:featuresSotA}
\begin{tabularx}{\columnwidth}{llX}
  {} & Feature & Description \\ \midrule  
 \multirow{3}{*}{\rotatebox[origin=c]{90}{\smaller{Size}}}  & LA & Number of lines added by the change \\
  & LD & Number of lines deleted by the change \\
  & LT & Number of lines in the files before the change  \\ \midrule   
  \multirow{4}{*}{\rotatebox[origin=c]{90}{\smaller{Diffusion}}}  & NS & Number of modified subsystems \\
  & ND & Number of modified directories \\
  & NF & Number of modified files \\
  & ENT & Spread of modified lines across modified files  \\ \midrule  
  \multirow{3}{*}{\rotatebox[origin=c]{90}{\smaller{History}}}  & NUC & Number of unique changes to all modified files  \\
  & NDEV & Number of past developers modifying the changed files \\
  & AGE & Time between last and current change to modified files \\ \midrule    
  \multirow{4}{*}[-1.7em]{\rotatebox[origin=c]{90}{Experience}} & EXP & Number of prior developer/reviewer changes \\
  & REXP & Number of prior developer changes weighted by the age of the respective change \\
  & SEXP & Number of prior changes to modified subsystems that an developer has participated in  \\
  & AWARE & Proportion of prior changes that an developer has participated in \\ \midrule  
  \multirow{4}{*}{\rotatebox[origin=c]{90}{\smaller{Review}}}  & ITER & Number of reviews that were performed \\
  & NREV & Number of persons, reviewing the changes \\
  & NCOM & Number of review comments \\
  & AGEREV & Time between the review request and the final approval \\ 
\end{tabularx}
\end{table}

\section{Approach} \label{Approach}

In this study, we investigate the effectiveness of different defect prediction features.
We have identified two new feature sets that were not yet utilized in JIT defect prediction research and may enhance the predictive power of models. 
In the following, we will describe the contained information and the intuition behind both feature sets. 

\subsection{State-Of-The-Art Features} \label{ApproachSOTA}

As shown in Section \ref{RelatedWorkRefs}, researchers have extensively studied which features are well suited for JIT fault prediction. We use these features as baseline for the subsequent evaluation of JIT defect prediction approaches.
All state-of-the-art features are referred to as \textit{SotA} in the following and listed in Table \ref{tab:featuresSotA}.
Since open-source development on GitHub mostly integrates features using pull requests, the embedded reviewing system is often used for code reviews. 
Developers can comment on code lines inside the pull request, leading to a more direct and integrated feedback.
Therefore, for extracting the review features from Kononenko et al. \cite{Kononenko2015}, we use data from the pull request review system and adjust the approach to reflect current software development practices. 
We believe, it is a fair adjustment and an appropriate way to enable the usage of well-understood existing defect prediction features in this context.

\subsection{Workflow Features} \label{ApproachI}

\begin{table}
\centering\small
 \caption{Software development workflow-based features for JIT defect prediction}
 \label{tab:featuresWORKFLOW}
\begin{tabularx}{\columnwidth}{llX}
   & Feature & Description \\ \midrule
\multirow{5}{*}{\rotatebox[origin=c]{90}{\smaller{Size}}}  
& DUR & Feature development time until the current change \\
 & NTH & Position of this change within the feature  \\
 & C-LA & Number of added lines in the feature \\
 & C-LD & Number of deleted lines in the feature  \\
 & SHARE & Share of current changes compared to whole feature development \\
  \midrule
 \multirow{4}{*}[-0.5em]{\rotatebox[origin=c]{90}{\smaller{Diffusion}}}   & C-NS & Number of modified subsystems in the feature \\
  & C-ND & Number of modified directories in the  feature  \\
  & C-NF & Number of modified files in the feature \\
  & C-ENT & Spread of modified lines across modified files in the feature \\ \midrule 
 \multirow{4}{*}{\rotatebox[origin=c]{90}{\smaller{Continuity}}}  & REL-NTH & Number of commits since last release \\
  & REL-DUR & Passed time since last release \\
  & DEV-NTH & Number of commits since last developer change \\
  & DEV-DUR & Passed time since last developer change \\\midrule 
 \multirow{2}{*}{\rotatebox[origin=c]{90}{\smaller{Comment}}} 
  & PRCOM & Number of comments  \\
  & PRCOM-R & Number of comment reactions \\
\end{tabularx}
\end{table}

\begin{figure}
    \centering
    \includegraphics[width=1.05\columnwidth]{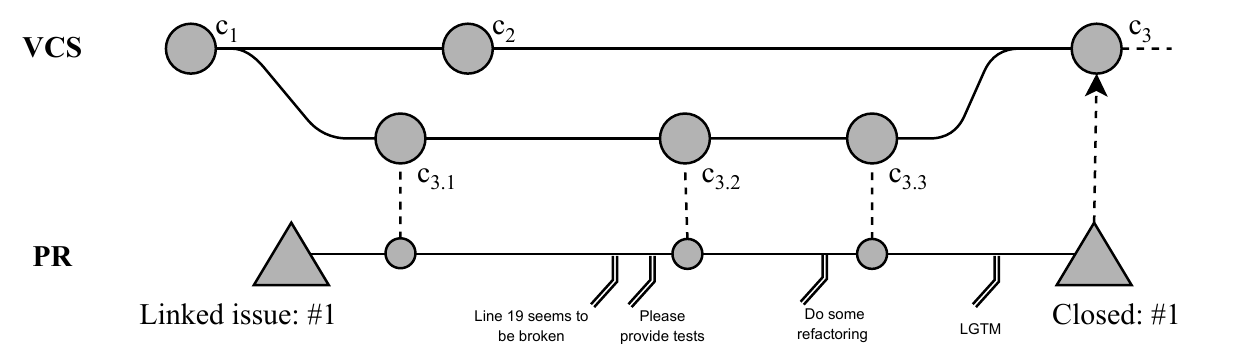}
    \caption{Schematic depiction of a feature development containing several inner pull request commits and comments.}
    \label{fig:workflow}
\end{figure}

Nowadays, open-source projects on GitHub are mostly developed applying a pull-based workflow, where contributors propose changes in a pull request and maintainers review, test and accept the changes if they find them suitable. These workflows and the corresponding pull requests contain information about the undertaken software development tasks and the complexity of changes. 
We depict an exemplary feature development graph in Figure \ref{fig:workflow}.
Feature development may concern additional functionality, maintenance activities or bug fixes.
Each feature consists of a pull request, a number of commits existing in the git history and potentially a list of commits that were present during feature development (inner-commits). In example, $c_{3.1-3}$ may be rebased, merged or squashed during pull request integration. Furthermore, the pull request contains information about feature development duration, progress, reviewers and review content, discussions and reactions.

While investigating workflow features for JIT defect prediction, we sort all data by creation timestamps to ensure temporal consistency.
For instance, if a commit is older than a comment in a pull request, this comment does not count towards the number of comments associated with this change.
In a rebased or squashed setting this is problematic since commits are recreated instead of merged ($c_{3.1-3}$ do not exist after integration) and all pull request information is associated with the new commit. 
We utilize information from internal feature development activities, reconstruct the development workflow and mitigate this problem. 
For this purpose, we map VCS commits to inner-commits by comparing hashes and commit messages and use the inner-commit whenever possible. 

We define 15 features listed in Table \ref{tab:featuresWORKFLOW} and denote them as \textit{WORKFLOW} features in the remainder of this study.
Similar, to the  added (LA) and removed (LD) lines from the SotA features, we define the cumulative counterparts (C-LA, C-LD) based on the position of the current change inside the pull request (NTH). 
Thus, these features represent not only the changes of the current commit, but are meant to reflect the entire feature development up to this point in time.
Additionally, we calculate the share of the current change compared to the previous feature development (SHARE) and the time spent for feature development up to the current change (DUR).
We do this to represent the status of feature development and to look at rapidly evolving or stale features. 

We do also report cumulative values for number of subsystems (C-NS), directories (C-ND), files (C-NF) and entropy (C-ENT) to include the full complexity of feature development.

In order to consider the fluctuating occurrence of defects over time and particularly after releases \cite{5463286}, we count the commits (REL-NTH) and measure the passed time (REL-DUR) since the last release. For our study, every commit associated with a git tag is marked as a release.
The same calculation is performed for the last commit since the developer's involvement (DEV-NTH, DEV-DUR) to represent recent developer activity and awareness.
Finally, we compute two features that represent community involvement and interest.

We count the number of comments (PRCOM) in the pull request before the current change, as well as the number of reactions from community member to these comments (PRCOM-R).

\subsection{AST Change Information Features} \label{ApproachII}

\begin{table}
\centering\small
 \caption{AST-based features for JIT defect prediction}
 \label{tab:featuresAST}
\begin{tabularx}{\columnwidth}{llX}
   & Feature  & Description \\ \midrule  
\multirow{6}{*}{\rotatebox[origin=c]{90}{\smaller{Methods}}}   & FUN & Number of functions in all files before the change \\
  & FUNT & Size of all functions before the change \\
  & FUNDIFF & Difference to FUNT after the change \\
  & FUNA & Number of added functions \\
  & FUND & Number of removed function \\
  & FUNU & Number of changed functions \\ \midrule  
  \multirow{12}{*}[2em]{\rotatebox[origin=c]{90}{\smaller{Nodes}}}   & ASTA & Number of added nodes to the AST \\
  & ASTD & Number of removed nodes from the source AST \\
  & ASTU & Number of updated nodes between both AST versions  \\
  & SASTA & Number of added special nodes to the AST \\
  & SASTD & Number of removed special nodes from the AST  \\
  & CASTA & Number of added comment nodes to the AST  \\
  & CASTD & Number of removed comment nodes from the AST  \\
  & PASTA & Number of added primitive nodes to the AST  \\
  & PASTD & Number of removed primitive nodes from the AST  \\ \midrule 
  \multirow{3}{*}{\rotatebox[origin=c]{90}{\smaller{Depth}}} & DEPTHA & Maximal depth of added nodes in the AST \\
  & DEPTHD & Maximal depth of removed nodes in the AST \\
  & DEPTHU & Maximal depth of updated nodes in the AST \\
\end{tabularx}
\end{table}

\begin{figure}
    \centering
    \includegraphics[width=.8\columnwidth]{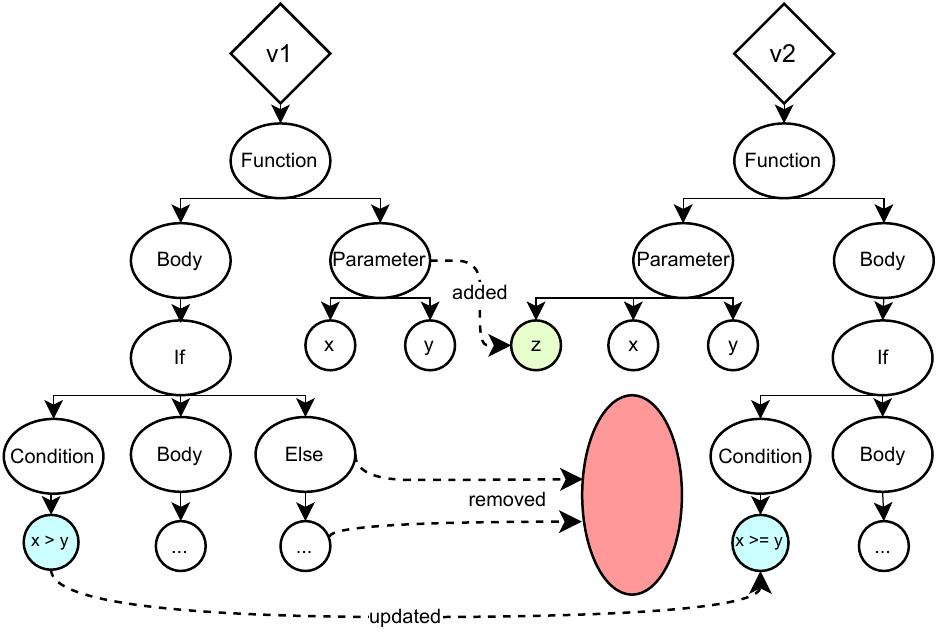}
    \vspace{-1em}
    \caption{Schematic illustration of changes in the AST.}
    \label{fig:AST-diff}
\end{figure}

Traditional JIT defect prediction approaches do not consider deeper information about the code change than number of changed lines and diffusion of the change. However, deep learning approaches do try to extract more syntactic and semantic information from the code change by embedding the change-log \cite{Hoang2019, Hoang2020} or the abstract syntax tree \cite{Dam2018}.
With JIT defect prediction, we do not only have information about the current state of the code but also about the change that happened between two versions of the source code. 
This change-level code information embeds knowledge about the complexity, depth and size of a change. 
Figure \ref{fig:AST-diff} shows a schematic representation of a AST change.
To our knowledge, such data has not been used in metrics-based JIT failure prediction approaches.
However, it may contribute to prediction effectiveness as it adds more detailed change information.

In the following, we define 18 features denoted as \textit{PATH} features in the remainder of this study and listed in Table \ref{tab:featuresAST}.
Similar to the length of a file, the number of methods can be an indicator of functionality without considering boilerplate code such as imports, class definitions or variable declarations.
We define FUN as the number of methods in all modified files. 
In a similar manner, we sum the lines of code in all methods before the change and denote them as FUNT. These two features are a measure of the size of the change reduced by code that does not contribute to functionality.
To measure how much functionality is changed we use FUNDIFF describing the line difference between the before- and after-version. 
We further count the number of added (FUNA), changed (FUNU) and deleted (FUND) methods. 
We argue that this also reflects refactorings (extracted code and methods) better than just recognizing the lines of a file.
Especially FUNU is of interest here, since the number of changed methods can be seen as a measure of change of functionality that is more prone to causing regression problems.

Second, we look at the changed nodes in the AST. We collect all added (ASTA), deleted (ASTD) and updated (ASTU) nodes, representing the change in more detail.
In contrast to added and removed lines, long lines could have several nodes that are added, removed or changed. 
An example of this can be seen in Figure \ref{fig:AST-diff}, where the added node does not result in a new line.
When looking at SotA features, this would result in one removed (LD) and one added line (LA), which may be too coarse to represent the change. 

For every changed node we classify the node as 'Primitive' (nodes representing strings, numbers, bytes, etc.), 'Comment' (nodes representing inline comments), 'Special' (nodes that represent control flow elements such as loops or if-clauses) or 'Default' (any node without a known special context). 
Thereby, we also report the number of added and deleted nodes in every group (SASTA, SASTD, CASTA, CASTD, PASTA, PASTD). Each of these groups has a different meaning and intuition.
Special instructions that regard the control flow of a method concern the cognitive complexity needed to understand the code, similar to the intuition behind cyclomatic complexity \cite{1702388}. Changes to inline comments might indicate more complex or unusual code \cite{martin2009clean}.
Changes to primitive code might hint at configuration or calculation changes. 

Lastly, we calculate the maximal depth of changes in the ASTs for every method (DEPTHA, DEPTHD, DEPTHU), using depth-first search.
The intuition is that deeper changes to the AST are more complex and therefore more error-prone.

\section{Study Design} \label{EmpiricalStudyRefs}

\begin{figure*}[htbp]
    \centering
    \includegraphics[width=\textwidth]{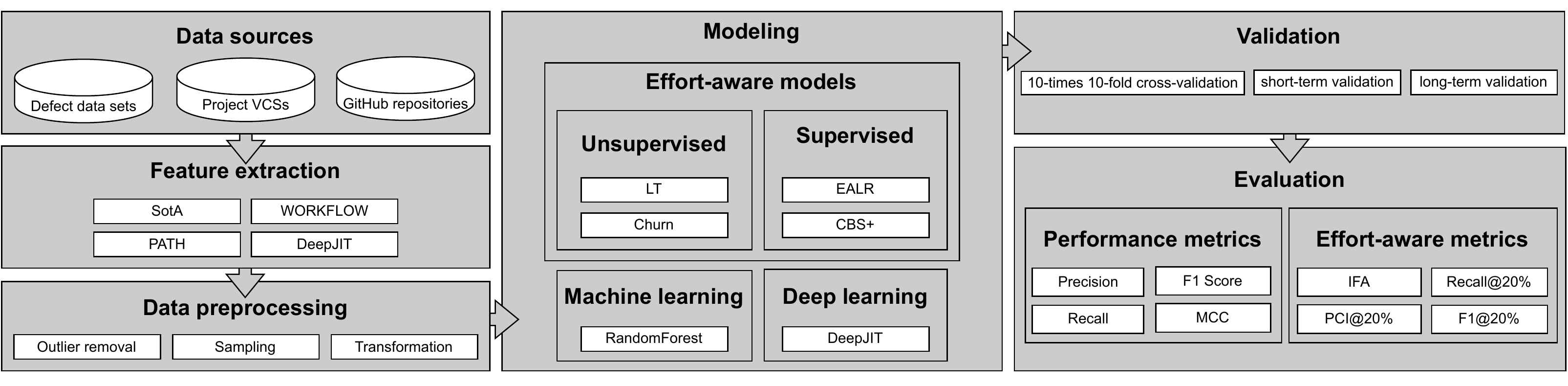}
    \vspace{-2em}
    \caption{Study overview and process.}
    \label{fig:study-overview}
\end{figure*}

In this study, we aim to investigate the effectiveness of metric-based JIT defect prediction. 
We gave an overview of state-of-the-art approaches and features as well as defined novel feature sets that potentially improve defect prediction.
In the remainder of this work, we investigate if these novel features do have an impact on JIT defect prediction effectiveness and which features are most relevant when applied to different project setups.  
Using these insights, we want to improve future JIT defect prediction applicability.

Figure \ref{fig:study-overview} shows an overview of the steps and evaluations we performed.
We present the evaluation setup for the conducted experiments in detail below to address the following research questions:

\begin{itemize}
    \item \textbf{RQ1:} How effective are the investigated features in terms of model performance?
    \item \textbf{RQ2:} How effective are the investigated features in terms of effort-awareness?
    \item \textbf{RQ3:} What are the most important features and do they differ between projects?
\end{itemize}

\paragraph{Data sources}  
For our study, we use a defect data set recently published by Bludau and Pretschner \cite{bludau2022}.
This data set consists of defect information from six large open-source projects with varying project characteristics and development workflows in the period from January 2015 to December 2020.
The data set is extracted applying a novel SZZ variant, that aims to identify defect-inducing commits utilizing the development history in git, information from  pull requests and issue management systems.
The projects and the data set characteristics are depicted in Table \ref{tab:projects}.
We further clone all six repositories and download GitHub repository information, such as pull request and inner commits, reviews and discussions.

\begin{table}
\vspace{-1em}
\caption{Defect data sets used in this study. Total number of commits is shown alongside number of distinct bug-inducing and -fixing commits for each project.}
\label{tab:projects}
\begin{center} \small
\begin{tabularx}{\columnwidth}{Xrrr}
\toprule
 Project & Commits & Inducing commits & Fixing commits \\
 \midrule
 Airflow &    7,311 &  1,582 (21.64\%) &  1,396 (19.09\%) \\
 Angular &   16,050 &  2,545 (15.86\%) &  2,013 (12.54\%)  \\
 Calcite &    2,463 &  774 (31.43\%) &   1,182 (47.99\%) \\
 Jenkins &    8,784 &  380 (4.33\%) &  750 (8.54\%)  \\
 Kafka &    5,812 &   1,377 (23.69\%) &   1,722 (29.63\%)  \\
 Pulsar &    4,127 &  1,016 (24.62\%) &    903 (21.88\%)  \\
 \bottomrule
\end{tabularx}
\end{center}
\vspace{-1.3em}
\end{table}

\paragraph{Feature extraction}  
For extracting SotA features, we apply the computations proposed by Kamei et al. \cite{Kamei2013a} and McIntosh et al. \cite{McIntosh2018a}.
We collect WORKFLOW features by using the retrieved pull request information.
We investigate every commit, look for a pull-request reference, find the inner-commits and calculate the features.
If a change does not have an associated pull request, then all related features are set to zero.
We collect the PATH features by parsing every changed file in the commit. We extract all methods using ANTLR grammars of the respective programming language.
We support, among others, the main languages used in the investigated projects, Python, Java, Javascript and Typescript.
For every changed method, we get the length before and after the change and inspect if it was added, removed or modified.
Additionally, we use GumTree \cite{gumtree} to compare the ASTs of both versions of the code and extract features accordingly.
Finally, we aggregate all changes of one commit.
To compare against deep learning approaches, we extract data to learn DeepJIT as proposed by Hoang et al. \cite{Hoang2019}.

\paragraph{Data preprocessing}
We neglect outliers (commits that change more than 10000 lines or 100 files) to keep our results consistent with previous authors \cite{McIntosh2018a}.
The number of ignored commits is relatively small ranging from 14 (Calcite) to 154 (Jenkins).
The data sets used in this study are highly imbalanced, since the majority of commits in a repository normally do not contain defects.
In the investigated projects, the proportion of defective changes ranges from 4.33\% (Jenkins) to 31.43\% (Calcite).
Therefore, we apply random down-sampling on the training data of all projects to mimic a balanced learning approach. This approach, used in previous studies \cite{Kamei2013a}, randomly removes non-defective commits (majority class) until the classes are balanced.
In fact, we tried different imbalance mitigation strategies and found that all of them produced similar results.
If different features are in different ranges, this could have an negative impact on the learning process \cite{589532}.
Therefore, we scale all features by removing the mean and scaling the data to unit variance based on the distribution of the testing data.
Moreover, most of the studied features are highly skewed.
As proposed by Kamei et al. \cite{Kamei2013a}, we apply a logarithmic transformation to the values of each non-binary feature to reduce data skewness. 

\paragraph{Modeling} \label{ModelTechRef}
We learn models with each of the described feature sets (SotA, WORKFLOW, PATH) and all features combined (ALL).
In our study, we learn Random Forest models as it is shown to perform well in JIT defect prediction \cite{Bowes2018, Osman2017}.
We use the standard implementation of the algorithm provided by \textit{scikit-learn}\footnote[1]{\url{https://scikit-learn.org/stable/}}. 
We additionally learn a state-of-the-art deep learning model, called DeepJIT, proposed by Hoang et al. \cite{Hoang2019}.
We prepare existing unsupervised and supervised effort-aware approaches described in Section \ref{RelatedWorkRefs}, to verify the validity of our evaluation results also in an effort-aware context.
We select two unsupervised approaches, denoted as CHURN \cite{Liu2017} and LT \cite{Yang2016} that solely sort the commits based on either the lines changed or by lines of code in the changed files before the change.
Additionally, we choose two supervised approaches, namely, EALR \cite{Kamei2013a} and CBS+ \cite{Huang2019c}, that are trained using the \textit{SotA} features.
As \textit{CBS+} is the best performing approach in literature, we apply the same sorting mechanism to our trained models with each feature set to make them effort-aware and evaluate them in this regard. 
Since JIT defect prediction is a binary classification task we apply a threshold to decide whether a change is potentially defective or not. For all our experiments, we set the threshold value to 0.5.
If the output of the model for a commit is above the threshold, the commit is classified as error-prone.

\paragraph{Validation} \label{SplitRef}

To evaluate the model performance, we divide the data sets in training and testing data. We employ three different validation settings described in the following. 
First, we perform a \textbf{10-times 10-fold cross validation} applying a stratified cross-validation approach to maintain the percentage of samples in each class. 
Second, McIntosh and Kamei \cite{McIntosh2018a} propose to validate JIT defect prediction models on a timely basis, since random sampling does not reflect the time-dependant nature of software development.
Due to restrictions from SZZ the last six months of development in the data sets contain relatively fewer commits labeled as defective.
We neglect them in our evaluation. 
In \textbf{short-term validation}, data from six months is used as training data to only consider the most recent development activities.
We split the data sets into six month time-frames. Beginning with the earliest frame ($n$) we use $n$ as training data and $n+1$ as test data.
We stop the process as soon as no more test data is available.
As more training data is not always available but has the potential to increase model performance, with \textbf{long-term validation} all historic data is used as training data.
Accordingly, we use the changes of the last six months as test data, while the rest of the data is used for training.
For the deep learning model we apply stratified a single 10-fold cross-validation.

\paragraph{Evaluation} \label{EvalRef}

We use common evaluation metrics to evaluate the model performance. 
For every model run, we calculate recall (R) and precision (P), based on the confusion matrix. We also calculate the f1-score (F1) representing the harmonic mean of recall and precision.
Furthermore, Matthews correlation coefficient (MCC) \cite{10.1093/bioinformatics/16.5.412} is calculated as it represents a measure that takes all four fields of the confusion matrix into account and is more reliable, especially when dealing with imbalanced data \cite{10.1145/3383219.3383232}.
For effort-aware performance, the percentage of found defects investigating only 20\% of changed lines, representing the effort-aware recall (R@20\%), is used in most studies (e,g., \cite{Yang2017, Liu2017, Yang2015}). 
Additionally, we calculate F1 at 20\% effort (F1@20\%), to not only report the completeness but also the correctness of the predictions.
Huang et al. \cite{Huang2019c} propose two metrics that we also investigate in our study. Accounting for the problem of frequent context switches, the proportion of commits that are reviewed when 20\% of all lines are inspected (PCI@20\%) is defined. 
To represent the induced developer frustration due to a lack of early wins, they define the number of false-positives before the first real defective commit is predicted by the model (IFA). For both metrics a smaller value is preferable.

\section{Results} \label{ResultsRef}

This section is organized based on the formulated research questions and discusses them accordingly. 
We provide all extracted features and necessary source code to replicate our evaluation as well as more detailed evaluation results in our supplementary material openly available on figshare \cite{Bludau2022SUPP}\footnote[2]{\url{https://doi.org/10.6084/m9.figshare.20199986}}.

\subsection{RQ1: How effective are the investigated features in terms of model performance?} \label{ResultsRefRQ1}

\begin{table*}
\centering\small
\caption{Performance of different feature sets in all three validation settings.}
\label{tab:RQ1_performance_scores}
\begin{tabularx}{\textwidth}{Xlrrrrrrrrrrrr}
\toprule
        \multicolumn{2}{l}{} & \multicolumn{4}{c}{\smaller{cross-validation}} & \multicolumn{4}{c}{\smaller{short-term validation}} & \multicolumn{4}{c}{\smaller{long-term validation}} \\
        Project & Feature set &  R & P & F1 &  MCC & R & P & F1 &  MCC & R & P & F1 &  MCC\\
\midrule
\multirow{4}{*}{Airflow} &       SotA &         0.72 & 0.42 &     0.53 & 0.37 &                  0.73 & 0.44 &     0.54 & 0.34 &                 0.89 & 0.33 &     0.48 & 0.32 \\
    &   WORKFLOW &                              0.71 & 0.40 &     0.51 & 0.35 &                  0.72 & 0.46 &     0.54 & 0.35 &                 0.81 & 0.34 &     0.48 & 0.31 \\
    &       PATH &                              0.70 & 0.40 &     0.50 & 0.34 &                  \textbf{0.76} & 0.45 &     0.56 & 0.37 &                 0.82 & \textbf{0.38}&     \textbf{0.52} & \textbf{0.38} \\
    &        ALL &                              \textbf{0.78 }& \textbf{0.47} &     \textbf{0.58} & \textbf{0.45} &                  0.75 & \textbf{0.52} &     \textbf{0.60} & \textbf{0.44} &                 \textbf{0.92} & 0.36 &     \textbf{0.52} & \textbf{0.38} \\\midrule
\multirow{4}{*}{Angular} &       SotA &         0.81 & 0.35 &     0.49 & 0.40 &                  0.69 & 0.33 &     0.44 & 0.32 &                 \textbf{0.77} & 0.36 &     0.49 & 0.39 \\
   &   WORKFLOW &                               0.75 & 0.31 &     0.44 & 0.33 &                  0.65 & 0.29 &     0.39 & 0.25 &                 0.71 & 0.33 &     0.45 & 0.32 \\
   &       PATH &                               0.69 & 0.36 &     0.47 & 0.36 &                  0.66 & \textbf{0.37} &     0.47 & 0.36 &                 0.48 & 0.42 &     0.45 & 0.34 \\
   &        ALL &                               \textbf{0.82 }& \textbf{0.36} &     \textbf{0.50 }& \textbf{0.42} &                  \textbf{0.73} & 0.36 &     \textbf{0.48} & \textbf{0.37} &                 0.72 & \textbf{0.42} &     \textbf{0.53} & \textbf{0.42 }\\\midrule
\multirow{4}{*}{Calcite} &       SotA &         0.74 & 0.54 &     0.63 & 0.43 &                  0.67 & 0.55 &     0.60 & 0.39 &                 0.59 & \textbf{0.49} &     0.54 & 0.37 \\
    &   WORKFLOW &                              0.62 & 0.43 &     0.50 & 0.22 &                  0.59 & 0.42 &     0.48 & 0.18 &                 0.63 & 0.33 &     0.43 & 0.17 \\
    &       PATH &                              \textbf{0.80} & 0.55 &     0.65 & 0.46 &                  \textbf{0.78} & 0.56 &     0.65 & 0.46 &                 \textbf{0.80} & 0.47 &     \textbf{0.59} & \textbf{0.44} \\
    &        ALL &                              \textbf{0.80} & \textbf{0.57} &     \textbf{0.66} & \textbf{0.49} &                  0.77 & \textbf{0.58} &     \textbf{0.66} & \textbf{0.48} &                 0.74 & 0.48 &     0.58 & 0.42 \\\midrule
\multirow{4}{*}{Jenkins} &       SotA &                              0.74 & 0.11 &     0.19 & 0.20 &                  0.56 & 0.08 &     0.14 & 0.12 &                 0.71 & 0.07 &     0.12 & 0.14 \\
 &   WORKFLOW &                              0.77 & 0.11 &     0.20 & 0.22 &                  0.70 & 0.09 &     0.16 & 0.17 &                 \textbf{0.73} & \textbf{0.09} &     \textbf{0.16} & 0.20 \\
 &       PATH &                              0.78 & 0.09 &     0.17 & 0.19 &                  0.69 & 0.09 &     0.15 & 0.16 &                 \textbf{0.73} & 0.06 &     0.12 & 0.15 \\
 &        ALL &                              \textbf{0.82} & \textbf{0.14} &     \textbf{0.24} & \textbf{0.28} &                  \textbf{0.77} & \textbf{0.13} &     \textbf{0.22} & \textbf{0.25} &                 \textbf{0.73} & \textbf{0.09} &     \textbf{0.16} & \textbf{0.21} \\\midrule
\multirow{4}{*}{Kafka} &       SotA &           0.74 & 0.48 &     0.58 & 0.43 &                  0.69 & 0.45 &     0.54 & 0.38 &                 0.66 & 0.36 &     0.47 & 0.34 \\
      &   WORKFLOW &                            0.72 & 0.45 &     0.55 & 0.39 &                  0.70 & 0.42 &     0.51 & 0.34 &                 0.56 & 0.39 &     0.46 & 0.33 \\
      &       PATH &                            0.87 & 0.52 &     0.65 & 0.54 &                  \textbf{0.86} & 0.51 &     0.64 & 0.53 &                 \textbf{0.92} & 0.36 &     0.52 & \textbf{0.44} \\
      &        ALL &                            \textbf{0.88} & \textbf{0.56} &     \textbf{0.69} & \textbf{0.59} &                  0.85 & \textbf{0.54} &     \textbf{0.65} & \textbf{0.55} &                 0.77 & \textbf{0.41} &     \textbf{0.53} & 0.43 \\\midrule
\multirow{4}{*}{Pulsar} &       SotA &          0.73 & 0.48 &     0.57 & 0.41 &                  0.68 & 0.47 &     0.56 & 0.38 &                 0.58 & 0.47 &     0.52 & 0.35 \\
     &   WORKFLOW &                             0.71 & 0.46 &     0.56 & 0.39 &                  0.76 & 0.41 &     0.53 & 0.33 &                 0.56 & 0.55 &     0.55 & 0.40 \\
     &       PATH &                             0.79 & 0.49 &     0.60 & 0.46 &                  \textbf{0.80} & 0.50 &     0.61 & 0.46 &                 \textbf{0.76} & 0.48 &     0.59 & 0.44 \\
     &        ALL &                             \textbf{0.80} & \textbf{0.51} &     \textbf{0.62} & \textbf{0.48} &                  \textbf{0.80} & \textbf{0.51} &     \textbf{0.62} & \textbf{0.47} &                 \textbf{0.76} & \textbf{0.50} &     \textbf{0.60} & \textbf{0.46} \\
\bottomrule
\end{tabularx}
\end{table*}

In order to answer \textit{RQ1}, Table \ref{tab:RQ1_performance_scores} shows the model performance for all feature sets and validation settings.
All projects but one show similar performance results in the range of 0.50 to 0.69 for F1 or 0.42 to 0.59 for MCC.
However, for the Jenkins project the performance is much lower in all validation settings and for every feature set. 
It becomes apparent that different projects benefit from features extracted from different sources. In example, the WORKFLOW features provide improvements in the Jenkins project, whereas Calcite benefits most from PATH features. 
In contrast, no novel feature set can increase the performance in Angular. However, the combination of all features (ALL) elevates the performance.
The performance is increased using all features in the majority of projects and validation settings.
Only little cases exist where single metrics show better results when trained on only one of the feature sets, e.g, the recall for Kafka in the long-term validation setting is higher using PATH features (0.92), than using ALL features (0.77).

\begin{table*}
\begin{center}\small
\caption{Significance and effect size of different feature sets in cross-validation for MCC on the left and R@20\% on the right.}
\label{tab:RQ1_performance_sig}
\begin{tabularx}{\textwidth}{Xclllclll}
\toprule
{} & \multicolumn{4}{c}{MCC} & \multicolumn{4}{c}{R@20\%} \\
{} & \multicolumn{1}{c}{SotA} &               \multicolumn{1}{c}{WORKFLOW} &                  \multicolumn{1}{c}{PATH} &                   \multicolumn{1}{c}{ALL} & \multicolumn{1}{c}{SotA} &              \multicolumn{1}{c}{WORKFLOW} &                   \multicolumn{1}{c}{PATH} &                   \multicolumn{1}{c}{ALL} \\
\midrule
Airflow    & 0.37 &   0.35 (-7.45\%)$^{****}_M$ &  0.34 (-8.76\%)$^{****}_L$ &  0.45 (21.27\%)$^{****}_L$ & 		0.58 &     0.59 (1.61\%) $^{ns}_N$ &    0.58 (-0.02\%) $^{ns}_N$ &  0.65 (13.67\%) $^{****}_L$ \\
Angular   & 0.40 &  0.33 (-17.33\%)$^{****}_L$ &  0.36 (-9.29\%)$^{****}_L$ &   0.42 (5.47\%)$^{****}_L$ & 			0.57 &  0.53 (-6.78\%) $^{****}_L$ &  0.52 (-8.67\%) $^{****}_L$ &   0.59 (4.92\%)$^{****}_M$ \\
Calcite    & 0.43 &  0.22 (-47.98\%)$^{****}_L$ &    0.46 (8.78\%)$^{***}_M$ &  0.49 (13.71\%)$^{****}_L$ & 		0.47 &  0.42 (-9.96\%) $^{****}_L$ &  0.52 (11.26\%) $^{****}_L$ &  0.52 (12.01\%) $^{****}_L$ \\
Jenkins & 0.20 & 0.22 (8.45\%)$^{*}_S$ &     0.19 (-8.76\%)$^{*}_S$ &  0.28 (35.48\%)$^{****}_L$ & 0.19 &      		0.71 (8.66\%) $^{****}_M$ &     0.67 (1.85\%) $^{ns}_N$ &  0.77 (18.17\%) $^{****}_L$ \\
Kafka      & 0.43 &    0.39 (-9.5\%)$^{****}_L$ &  0.54 (25.29\%)$^{****}_L$ &  0.59 (35.66\%)$^{****}_L$ & 		0.45 &    0.44 (-2.61\%) $^{ns}_S$ &   0.6 (32.39\%) $^{****}_L$ &  0.61 (35.24\%) $^{****}_L$ \\
Pulsar     & 0.41 &   0.39 (-6.97\%)$^{****}_M$ &  0.46 (10.42\%)$^{****}_L$ &  0.48 (15.54\%)$^{****}_L$ & 		0.50 &  0.47 (-5.78\%) $^{****}_M$ &  0.56 (10.78\%) $^{****}_L$ &  0.56 (12.01\%)$^{****}_L$ \\ \midrule
Average             & 0.38 &      0.32 (-13.46\%)$_L$ &       0.39 (2.95\%)$_L$ &      0.45 (21.19\%)$_L$ & 		0.54 &      0.53 (-2.47\%)$_M$ &       0.57 (7.93\%)$_L$ &       0.62 (16.0\%)$_L$ \\
\bottomrule
\end{tabularx}
\end{center}
\footnotesize 
\raggedleft
\emph{Significance levels:} p < 1e-04: ****, p < 0.001: ***, p < 0.01: **, p < 0.1: *, p >= 0.1: ns
\hfill\raggedright
\emph{Effect size classes:} $\delta$ < 0.147: 'N', $\delta$ < 0.33: 'S', $\delta$ < 0.474: 'M', $\delta$ >= 0.474: 'L'
\end{table*}

We investigate whether the differences in performance between \textit{SotA} and respective other feature sets are statistically significant.
In accordance with other studies \cite{Yan2020, Yang2017, Yan2019}, the Wilcoxon signed-rank test \cite{10.2307/3001968} at a 95\% significance level with Bonferroni correction \cite{Bonferonni} is applied to compare the feature sets against \textit{SotA} using all 100 runs from cross validation. 
The Bonferroni correction mitigates the problem of multiple comparisons. 
Additionally, we employ Cliff’s delta ($\delta$) \cite{cliff1996ordinal} to measure the effect size of the difference between the feature sets.
We use the effect-size mapping, proposed by Cliff, rendering a $\delta$-value below 0.147 'negligible', between 0.147 and 0.33 'small', between 0.33 and 0.474 'medium' and above 0.474 'large'. 
Results for MCC are shown on the left side of Table \ref{tab:RQ1_performance_sig}.
The models using all features improve the MCC measure on average significantly by 7 percentage points (pp) with a large effect size.  
The WORKFLOW features do not contribute much to the overall effectiveness of JIT defect prediction when applied alone. The feature set improves MCC only in the Jenkins project, but has a small effect size. The PATH features improve the model performance in three out of six investigated projects significantly with a medium to large effect size. 
We evaluated significance of improvement for every performance metric and validation setting. 
The combination of features does have an significant effect on the performance for every other performance metric, in a similar way.
Compared to the SotA features the improvements in the cross/short-term/long-term validation settings are on average 
12.28\%/20.99\%/13.96\% for P, 
9.73\%/16.46\%/12.14\% for R,  
12.12\%/20.83\%/15.25\% for F1 and
21.19\%/42.32\%/24.61\% for MCC.
We conclude that the combination of feature sets and sources of information can improve model effectiveness over the improvements that features of a single source can provide alone.
Detailed performance evaluation are provided in the supplementary material.

Interestingly, the performance improvement with respect to each performance measure is higher in the time-dependent validation than when cross-validation is used.
This may indicate that the predictive power of the different feature sets shifts over the life of the project and that no feature set performs best over the entire project life cycle.
When looking at the individual validation runs (100 for every project in cross-, 7 runs in short-term- and 1 run in long-term-validation), the results from the cross-validation are quite stable with a mean standard deviation of 0.03 over all projects. However, when using short-term models the standard deviation is higher (0.06) and varying prediction performance can be recognized between the six-month time frames. 
This shows, that different features from different sources are more important depending on the current software development activities and status of the project. 
Please see our supplementary material for further detail.

We conclude that features from different sources increase the performance and should be applied in JIT defect prediction. 
The combination of different features from several sources increases the performance over the effects one single feature source can provide.
Therefore, features from various sources are beneficial and should be combined to produce more effective JIT defect prediction models.

\vspace{.15cm}
\noindent\fbox{\parbox{.97\columnwidth}{\textit{
Combining SotA with novel workflow and AST change features leads to the best performance in all projects. 
In short-term validation, the performance is dependant on varying features over time. This should be considered and models should be frequently retrained.
}}}
\vspace{.1cm}

\subsection{RQ2: How effective are the investigated features in terms of effort-awareness?} \label{ResultsRefRQ2}

In order to answer \textit{RQ2}, we perform the same analysis as in \textit{RQ1} for effort-aware measures.
A counterpart to Table \ref{tab:RQ1_performance_scores} is contained in the supplementary material for effort-aware performance measures.
The same picture as in \textit{RQ1} emerges when investigating the effectiveness of each feature set to effort-awareness. 
The combination of all features is the best performing variation in terms of IFA, PCI@20\%, R@20\% and F1@20\% across all projects and validation settings. 
Considering only R@20\%, unsupervised existing approaches are often comparable to more elaborate supervised models.
Considering F1@20, IFA and PCI@20, the models trained with the combined feature set (ALL) and applying classification and sorting (CBS+) provides the best results in all evaluation settings.

We evaluated the significance and effect size of every effort-aware performance metric considered in this study, in the same way as presented in \textit{RQ1}.
An evaluation for R@20\% can be seen on the right of Table \ref{tab:RQ1_performance_sig}.
Compared to the SotA R@20\% is increased by 16.00\% (\SI{8}{pp}).
Using all features, every performance metric, except PCI@20\% is significantly improved in every validation setting. 
Compared to SotA the improvements in the cross/short-term/long-term validation settings are on average 
23.51\%/35.13\%/70.24\% for IFA, 
-1.85\%/-0.99\%/-4.23\% for PCI@20\%,
16.00\%/23.38\%/19.23\% for R@20\% and
15.95\%/24.12\%/20.28\% for F1@20\%.

\begin{table}
\centering\small
\caption{R@20\% in different model set-ups.}
\label{tab:RQ1_performance_sig_model}
\begin{tabular}{lrll}
\toprule
{} &  \multicolumn{1}{c}{CBS+$_{ALL}$} &                                     \multicolumn{1}{c}{Churn} &                  \multicolumn{1}{c}{CBS+$_{SotA}$}  \\
\midrule
Airflow    & 0.65  &     0.64 (1.81\%)$^{ns}_N$ &  0.53 (24.05\%)$^{****}_L$ \\
Angular   & 0.59 &   0.52 (15.4\%)$^{****}_L$ &   0.5 (18.73\%)$^{****}_L$ \\
Calcite    & 0.52  &  0.47 (11.18\%)$^{****}_L$ &  0.45 (15.52\%)$^{****}_L$ \\
Jenkins & 0.77  &   0.6 (28.64\%)$^{****}_L$ &   0.58 (32.3\%)$^{****}_L$ \\
Kafka      & 0.61  &  0.48 (28.66\%)$^{****}_L$ &  0.44 (38.85\%)$^{****}_L$ \\
Pulsar     & 0.56  &    0.56 (-0.32\%)$^{ns}_N$ &  0.46 (23.73\%)$^{****}_L$ \\ \midrule
Average    & 0.62 &      0.54 (14.23\%)$_L$ &      0.49 (25.53\%)$_L$ \\
\bottomrule
\vspace{-2em}
\end{tabular}
\end{table}

In addition, we apply the Wilcoxon signed-rank test \cite{10.2307/3001968} at a 95\% significance level with Bonferroni correction \cite{Bonferonni}, comparing the CBS+$_{ALL}$, learned with all features, to existing unsupervised and supervised effort-aware models.
In Table \ref{tab:RQ1_performance_sig_model}, we show a exemplary evaluation for effort-aware recall.
R@20\% is significantly increased by 14.23\% (\SI{8}{pp}) in respect to the best unsupervised and by 25.53\% (\SI{13}{pp}) for CBS+$_{SotA}$.
The results are significant for most projects showing a large effect size.
This is also true for other evaluation metrics, such as F1@20\% (107.48\%/28.08\%), or in short-term validation (8.14\%/21.51\%).
A trade-off between R@20\% of effort-aware models and developer usability becomes apparent.
Looking at R@20\% unsupervised models can have a comparable performance with supervised models and can even outperform them. The new feature sets elevate the performance which increases the difference.
However, investigating also the correctness (F1@20\%) and usability (IFA), supervised models have a large positive effect. 
As precision of JIT defect prediction models has a huge impact on developer acceptance, we argue that such metrics should be investigated more closely in future studies.

\vspace{.15cm}
\noindent\fbox{\parbox{.97\columnwidth}{\textit{
The additional features improve the performance significantly also in an effort-aware context.
Supervised models that use feature combination and subsequent sorting (CBS+) perform best.
Considering not only R@20\% but also the precision of predictions, CBS+ using an combination of all features always outperforms existing approaches.
}}}
\vspace{.1cm}

\subsection{RQ3: What are the most important features and do they differ between projects?} \label{ResultsRefRQ3}

In order to answer \textit{RQ3}, we conduct a feature importance analysis.
We perform a correlation analysis on all features and calculate feature importance using the mean decrease impurity for each feature.
We remove features for every project based on importance scores. 
We want to point out that some features are removed in all projects (e.g., NCOM).
Then, we learn Random Forest models again using the reduced feature sets for every project. 
We apply 10-times 10-fold cross validation, since it better represents features learned over the whole distribution of available data.
Therefore, 100 model learning runs are executed, leading to 100 importance scores for every feature.
We execute the Non-Parametric ScottKnott ESD (NPSK) test \cite{scottknott} to determine which features are the most important ones in the reduced set, similar to previous studies \cite{Yan2020, Fan2020, Yan2019}.
The NPSK test relaxes assumptions from the original ScottKnott test \cite{10.2307/2529204} in the sense that it does not assume normal distributed data or homogeneous distributions. 
It partitions all features into groups, maximizing the median value of the feature distributions between groups and merges groups that have a negligible effect size based on Cliff’s delta. 
Table \ref{tab:RQ3_feature_importance_project} shows the most important (Top-3) features for every project. 

\begin{table}
\centering\small
\caption{Most important (Top-3) features by project.}
\label{tab:RQ3_feature_importance_project}
\begin{tabularx}{\columnwidth}{lcccccc}
\toprule
        & \small{Airflow} & \small{Angular} & \small{Calcite} & \small{Jenkins} & \small{Kafka} & \small{Pulsar} \\ \midrule
AGE     & 1                                & -                                & -                                & -                                & -                              & -                               \\
ASTA    & 2                                & 2                                & 1                                & 3                                & 1                              & 1                               \\
C-LA    & 1                                & 2                                & -                                & 1                                & -                              & -                               \\
DEPTHA  & 3                                & -                                & 1                                & 3                                & 2                              & 2                               \\
ENT     & -                                & 2                                & -                                & -                                & -                              & -                               \\
FUN     & -                                & -                                & 2                                & 2                                & 3                              & -                               \\
FUNDIFF & -                                & -                                & -                                & -                                & -                              & 3                               \\
FUNT    & -                                & 2                                & -                                & -                                & -                              & -                               \\
LA      & -                                & 3                                & -                                & -                                & -                              & -                               \\
LT      & -                                & 1                                & 3                                & 2                                & 2                              & -                               \\
PASTA   & -                                & -                                & -                                & -                                & -                              & 3                               \\
SASTA   & -                                & -                                & -                                & -                                & -                              & 2                               \\ 
\bottomrule
\end{tabularx}
\footnotesize 
\raggedright
The '-' symbol represents that a feature was not in the top-3 ranked group for the respective project. Features that were not among the Top-3 for any project are not listed.
\vspace{-1.3em}
\end{table}

We confirm the findings from \textit{RQ1} that the PATH variables seem to have an important influence on model performance in general. 
As shown in Table \ref{tab:RQ3_feature_importance_project}, ASTA is the most important feature for three and under the top-3 most important feature for the other projects.
DEPTHA is among the top-3 for five projects, and selected in the top-5 for the other two.
Also, some WORKFLOW (e.g., C-LA) and SoTA features (e.g., LT) are highly ranked and frequently selected as most important features. 
This shows that all the different feature sets contribute to some extent to the overall prediction performance.

Project dependent differences become apparent when looking at Table \ref{tab:RQ3_feature_importance_project}. 
For instance, Angular has most important features (LA, ENT, FUNT) that are not in the top selected features for any other project. 
Some features are not among the most important features in any project, nor do they have a major impact on the model's performance.
However, there is no obvious way to differentiate which features will perform well in new project contexts.
In future, we want to investigate the relationship between features and different project characteristics to determine which sources of information should be considered in JIT defect prediction for a specific project.

\vspace{.15cm}
\noindent\fbox{\parbox{.97\columnwidth}{\textit{
Features from the PATH feature set (ASTA, DEPTHA) have an high influence on identifying defective commits for all investigated projects.
However, the varying importance of features is evident with different project environments.
}}}
\vspace{.1cm}

\section{Discussion} \label{Discussion}

In this section, we want to discuss some further findings and considerations that came up during the execution of our study.

\vspace{-.6em} 
\paragraph{Different classifiers} \label{DiffClassifierRef}

Throughout our study we report results for Random Forest models.
Nonetheless, we did evaluate a collection of other machine learning (Logistic Regression, AdaBoost, Multi-layer Perceptron classifier) and include their performance evaluation in our supplementary material for the sake of completeness. 
In line with other research, Random Forest is the best performing model regarding prediction effectiveness and effort-awareness across all projects. 
Considering P or PCI@20, other models perform better with the data sets used in this study, however, the differences are marginal and not significant
(P in 2 of 6 with a maximum of \SI{1}{pp} difference and PCI@20 in 3 of 6 projects with a maximum of \SI{2}{pp} difference, respectively). 
Since training is relatively simple and cheap, it is advisable to test different classifiers and select them according to important performance metrics in the particular context.

\vspace{-.6em} 
\paragraph{Novel feature sets}

We have shown that novel features can support the performance of JIT defect prediction models. 
The best performing features for one project are, however, not necessarily from the same feature set.
The benefit comes from mixing different kind of features extracted from different sources of the software and/or development process. 
However, the extraction of each feature comes with additional effort, since the data needs to be gathered, processed and accumulated.
With the results shown above, we often see that the WORKFLOW features do not elevate performance for most projects.
However, one cannot be sure without testing the features in the current project context, as can be seen especially in the Jenkins project. 
As our study also tested different classifiers and investigated the feature importance, we conclude that feature preparation and selection is more important than model selection.

\vspace{-.6em} 
\paragraph{Deep learning}

We also compare the model results to a deep learning approach to investigate the finding from recent studies \cite{Fu2017a, Zeng2021}, showing that deep learning does not always outperform metric-based approaches and is often even worse.
In line with their results, deep learning does not increase the performance of JIT defect prediction on the six investigated projects, in this study.
Applying a state-of-the-art deep learning approach leads to worse prediction results in all projects and all but one evaluation metrics. 
The deep learning approach outperforms the best metric-based approach in terms of R@20\% on average by 10\%. 
This means that DeepJIT finds more true defects when investigating only 20\% of lines.
However, these result may be misleading as F1@20\% shows worse performance in all projects. 
DeepJIT can find more true defective commits with less effort, but does also predict more incorrect commits as error-prone, which may lead to developer frustration. 
More information on our evaluation is provided in the supplementary material.
We conclude researchers should evaluate their deep learning approaches also against state-of-the-art metric-based JIT defect prediction approaches as presented throughout this study.

\vspace{-.4em}
\section{Threats to Validity} \label{ThreatsToValidityRefs}

Threats to \textbf{internal validity} refer to errors in our implementation and in the feature extraction phase for every commit.
To mitigate the risks of own coding errors, we use well known open-source third-party libraries and tools. 
For instance, we apply ANTLR grammars to find methods imports and comments inside code files.
We use GumTree to differentiate ASTs and extract change paths. 
In addition, to reduce the threats in our implementation, we double-checked and tested the code, but there could still exist problems we did not find.
The data sets used in this study are extracted using SZZ. Research has shown limitations, which may lead to wrongly labeled data.
We use a recently published data set of six open-source projects
created with a novel evaluated variant of the SZZ algorithm, generating more accurate data sets, 
to minimize the risk. 

Threats to \textbf{external validity} relate to the generalizability of our results.
We use six open source projects, which belong to different application domains, vary in size, cover five years of development activity and are written in different programming languages.
However, further studies analyzing more projects are needed to reduce this threat.
Since all projects are open source, no statement can be made about transferability pf the results to industrial projects.

Threats to \textbf{construct validity} refer to the soundness and suitability of our study design and evaluation.
We consider 10-times 10-fold cross validation to mitigate sampling bias, as well as, time-dependent validation of models to be more realistic. 
We evaluate all scenarios and compare their results in order to be more complete.
We apply different statistical tests to evaluate the significance of features in the context of JIT defect prediction.
All of the measures and tests are well-established and have been used in past studies.

\vspace{-.4em}
\section{Conclusion} \label{ConclusionRefs}

In this study, we investigate the performance of state-of-the-art features for JIT defect prediction.
Additionally, we follow a call from other studies and propose two novel feature sets based on the software development workflow and abstract syntax tree changes.
We perform an empirical evaluation and compare the effectiveness of the various features in JIT defect prediction on six open-source projects.
Our novel features do outperform state-of-the-art approaches in terms of standard and effort-aware performance metrics.
Models show varying performance dependent on the used features, however the combination of features improves the prediction performance in all cases.
For instance, the performance in terms of MCC is increased by 21.19\% (\SI{7}{pp}), on average. 
Evaluating effort-awareness, the recall investigating 20\% of changed lines is improved by 14.23\% (\SI{8}{pp}), on average.
We conclude, that novel features improve metric-based JIT defect prediction and often outperform even more sophisticated deep learning approaches. 

Added nodes and their depth in the AST are among the top-3 most important features to determine defective commits for all but one investigated projects. 
Apart from that, the most important features do vary based on the investigated projects and can be from any of the proposed feature sets.
Features and used data sources must be selected with care, to include the most differentiating features.

\begin{acks}
This study is based upon work supported by the \textit{Bavarian Ministry of Economic Affairs, Regional Development and Energy} under the \textit{Center for Code Excellence} project.
\end{acks}

\bibliographystyle{ACM-Reference-Format}
\bibliography{main}


\begin{thebibliography}{48}


\ifx \showCODEN    \undefined \def \showCODEN     #1{\unskip}     \fi
\ifx \showDOI      \undefined \def \showDOI       #1{#1}\fi
\ifx \showISBNx    \undefined \def \showISBNx     #1{\unskip}     \fi
\ifx \showISBNxiii \undefined \def \showISBNxiii  #1{\unskip}     \fi
\ifx \showISSN     \undefined \def \showISSN      #1{\unskip}     \fi
\ifx \showLCCN     \undefined \def \showLCCN      #1{\unskip}     \fi
\ifx \shownote     \undefined \def \shownote      #1{#1}          \fi
\ifx \showarticletitle \undefined \def \showarticletitle #1{#1}   \fi
\ifx \showURL      \undefined \def \showURL       {\relax}        \fi
\providecommand\bibfield[2]{#2}
\providecommand\bibinfo[2]{#2}
\providecommand\natexlab[1]{#1}
\providecommand\showeprint[2][]{arXiv:#2}

\bibitem[Abdi(2007)]%
        {Bonferonni}
\bibfield{author}{\bibinfo{person}{Hervé Abdi}.}
  \bibinfo{year}{2007}\natexlab{}.
\newblock \showarticletitle{The Bonferonni and Šidák Corrections for Multiple
  Comparisons}.
\newblock \bibinfo{journal}{\emph{Encyclopedia of measurement and statistics}}
  \bibinfo{volume}{3} (\bibinfo{date}{01} \bibinfo{year}{2007}).
\newblock


\bibitem[Arisholm et~al\mbox{.}(2010)]%
        {Arisholm2010}
\bibfield{author}{\bibinfo{person}{Erik Arisholm}, \bibinfo{person}{Lionel~C.
  Briand}, {and} \bibinfo{person}{Eivind~B. Johannessen}.}
  \bibinfo{year}{2010}\natexlab{}.
\newblock \showarticletitle{A systematic and comprehensive investigation of
  methods to build and evaluate fault prediction models}.
\newblock \bibinfo{journal}{\emph{Journal of Systems and Software}}
  \bibinfo{volume}{83}, \bibinfo{number}{1} (\bibinfo{date}{Jan.}
  \bibinfo{year}{2010}), \bibinfo{pages}{2--17}.
\newblock


\bibitem[Aversano et~al\mbox{.}(2007)]%
        {Aversano2007}
\bibfield{author}{\bibinfo{person}{Lerina Aversano}, \bibinfo{person}{Luigi
  Cerulo}, {and} \bibinfo{person}{Concettina~Del Grosso}.}
  \bibinfo{year}{2007}\natexlab{}.
\newblock \showarticletitle{Learning from bug-introducing changes to prevent
  fault prone code}. In \bibinfo{booktitle}{\emph{Ninth international workshop
  on Principles of software evolution in conjunction with the 6th {ESEC}/{FSE}
  joint meeting - {IWPSE} {\textquotesingle}07}}. \bibinfo{publisher}{{ACM}
  Press}.
\newblock
\urldef\tempurl%
\url{https://doi.org/10.1145/1294948.1294954}
\showDOI{\tempurl}


\bibitem[Bachmann and Bernstein(2010)]%
        {5463286}
\bibfield{author}{\bibinfo{person}{Adrian Bachmann} {and}
  \bibinfo{person}{Abraham Bernstein}.} \bibinfo{year}{2010}\natexlab{}.
\newblock \showarticletitle{When process data quality affects the number of
  bugs: Correlations in software engineering datasets}. In
  \bibinfo{booktitle}{\emph{2010 7th IEEE Working Conference on Mining Software
  Repositories (MSR 2010)}}. \bibinfo{publisher}{{IEEE}},
  \bibinfo{pages}{62--71}.
\newblock
\urldef\tempurl%
\url{https://doi.org/10.1109/MSR.2010.5463286}
\showDOI{\tempurl}


\bibitem[Baldi et~al\mbox{.}(2000)]%
        {10.1093/bioinformatics/16.5.412}
\bibfield{author}{\bibinfo{person}{Pierre Baldi}, \bibinfo{person}{Søren
  Brunak}, \bibinfo{person}{Yves Chauvin}, \bibinfo{person}{Claus A.~F.
  Andersen}, {and} \bibinfo{person}{Henrik Nielsen}.}
  \bibinfo{year}{2000}\natexlab{}.
\newblock \showarticletitle{{Assessing the accuracy of prediction algorithms
  for classification: an overview }}.
\newblock \bibinfo{journal}{\emph{Bioinformatics}} \bibinfo{volume}{16},
  \bibinfo{number}{5} (\bibinfo{date}{05} \bibinfo{year}{2000}),
  \bibinfo{pages}{412--424}.
\newblock
\showISSN{1367-4803}


\bibitem[Bludau and Pretschner(2022a)]%
        {Bludau2022SUPP}
\bibfield{author}{\bibinfo{person}{Peter Bludau} {and}
  \bibinfo{person}{Alexander Pretschner}.} \bibinfo{year}{2022}\natexlab{a}.
\newblock \showarticletitle{{Feature sets in just-in-time defect prediction}}.
\newblock  (\bibinfo{date}{8} \bibinfo{year}{2022}).
\newblock
\urldef\tempurl%
\url{https://doi.org/10.6084/m9.figshare.20199986.v1}
\showDOI{\tempurl}


\bibitem[Bludau and Pretschner(2022b)]%
        {bludau2022}
\bibfield{author}{\bibinfo{person}{Peter Bludau} {and}
  \bibinfo{person}{Alexander Pretschner}.} \bibinfo{year}{2022}\natexlab{b}.
\newblock \showarticletitle{PR-SZZ: How pull requests can support the tracing
  of defects in software repositories}. In \bibinfo{booktitle}{\emph{2022 IEEE
  International Conference on Software Analysis, Evolution and Reengineering
  (SANER)}}. \bibinfo{pages}{1--12}.
\newblock
\urldef\tempurl%
\url{https://doi.org/10.1109/SANER53432.2022.00012}
\showDOI{\tempurl}


\bibitem[Bowes et~al\mbox{.}(2017)]%
        {Bowes2018}
\bibfield{author}{\bibinfo{person}{David Bowes}, \bibinfo{person}{Tracy Hall},
  {and} \bibinfo{person}{Jean Petri{\'{c}}}.} \bibinfo{year}{2017}\natexlab{}.
\newblock \showarticletitle{Software defect prediction: do different
  classifiers find the same defects?}
\newblock \bibinfo{journal}{\emph{Software Quality Journal}}
  \bibinfo{volume}{26}, \bibinfo{number}{2} (\bibinfo{date}{Feb.}
  \bibinfo{year}{2017}), \bibinfo{pages}{525--552}.
\newblock
\urldef\tempurl%
\url{https://doi.org/10.1007/s11219-016-9353-3}
\showDOI{\tempurl}


\bibitem[Cliff(1996)]%
        {cliff1996ordinal}
\bibfield{author}{\bibinfo{person}{N. Cliff}.} \bibinfo{year}{1996}\natexlab{}.
\newblock \bibinfo{booktitle}{\emph{Ordinal Methods for Behavioral Data
  Analysis}}.
\newblock \bibinfo{publisher}{Erlbaum}.
\newblock
\showISBNx{9780805813333}
\showLCCN{96022689}


\bibitem[Dam et~al\mbox{.}(2018)]%
        {Dam2018}
\bibfield{author}{\bibinfo{person}{Khanh~Hoa Dam}, \bibinfo{person}{Trang
  Pham}, \bibinfo{person}{Shien~Wee Ng}, \bibinfo{person}{T. Tran},
  \bibinfo{person}{John~C. Grundy}, \bibinfo{person}{Aditya~K. Ghose},
  \bibinfo{person}{Taeksu Kim}, {and} \bibinfo{person}{Chul-Joo Kim}.}
  \bibinfo{year}{2018}\natexlab{}.
\newblock \showarticletitle{A deep tree-based model for software defect
  prediction}.
\newblock \bibinfo{journal}{\emph{ArXiv}}  \bibinfo{volume}{abs/1802.00921}
  (\bibinfo{year}{2018}).
\newblock


\bibitem[Eyolfson et~al\mbox{.}(2011)]%
        {Eyolfson2011}
\bibfield{author}{\bibinfo{person}{Jon Eyolfson}, \bibinfo{person}{Lin Tan},
  {and} \bibinfo{person}{Patrick Lam}.} \bibinfo{year}{2011}\natexlab{}.
\newblock \showarticletitle{Do time of day and developer experience affect
  commit bugginess}. In \bibinfo{booktitle}{\emph{Proceeding of the 8th working
  conference on Mining software repositories - {MSR} {\textquotesingle}11}}.
  \bibinfo{publisher}{{ACM} Press}.
\newblock


\bibitem[Falleri et~al\mbox{.}(2014)]%
        {gumtree}
\bibfield{author}{\bibinfo{person}{Jean{-}R{\'{e}}my Falleri},
  \bibinfo{person}{Flor{\'{e}}al Morandat}, \bibinfo{person}{Xavier Blanc},
  \bibinfo{person}{Matias Martinez}, {and} \bibinfo{person}{Martin Monperrus}.}
  \bibinfo{year}{2014}\natexlab{}.
\newblock \showarticletitle{Fine-grained and accurate source code
  differencing}. In \bibinfo{booktitle}{\emph{{ACM/IEEE} International
  Conference on Automated Software Engineering, {ASE} '14, Vasteras, Sweden -
  September 15 - 19, 2014}}. \bibinfo{pages}{313--324}.
\newblock


\bibitem[Fan et~al\mbox{.}(2020)]%
        {Fan2020}
\bibfield{author}{\bibinfo{person}{Yuanrui Fan}, \bibinfo{person}{Xin Xia},
  \bibinfo{person}{David Lo}, {and} \bibinfo{person}{Ahmed~E. Hassan}.}
  \bibinfo{year}{2020}\natexlab{}.
\newblock \showarticletitle{Chaff from the Wheat: Characterizing and
  Determining Valid Bug Reports}.
\newblock \bibinfo{journal}{\emph{{IEEE} Transactions on Software Engineering}}
  \bibinfo{volume}{46}, \bibinfo{number}{5} (\bibinfo{date}{May}
  \bibinfo{year}{2020}), \bibinfo{pages}{495--525}.
\newblock


\bibitem[Fu and Menzies(2017a)]%
        {Fu2017a}
\bibfield{author}{\bibinfo{person}{Wei Fu} {and} \bibinfo{person}{Tim
  Menzies}.} \bibinfo{year}{2017}\natexlab{a}.
\newblock \showarticletitle{Easy over hard: a case study on deep learning}. In
  \bibinfo{booktitle}{\emph{Proceedings of the 2017 11th Joint Meeting on
  Foundations of Software Engineering}}. \bibinfo{publisher}{{ACM}}.
\newblock
\urldef\tempurl%
\url{https://doi.org/10.1145/3106237.3106256}
\showDOI{\tempurl}


\bibitem[Fu and Menzies(2017b)]%
        {Fu2017}
\bibfield{author}{\bibinfo{person}{Wei Fu} {and} \bibinfo{person}{Tim
  Menzies}.} \bibinfo{year}{2017}\natexlab{b}.
\newblock \showarticletitle{Revisiting unsupervised learning for defect
  prediction}. In \bibinfo{booktitle}{\emph{Proceedings of the 2017 11th Joint
  Meeting on Foundations of Software Engineering}}. \bibinfo{publisher}{{ACM}}.
\newblock
\urldef\tempurl%
\url{https://doi.org/10.1145/3106237.3106257}
\showDOI{\tempurl}


\bibitem[Hoang et~al\mbox{.}(2019)]%
        {Hoang2019}
\bibfield{author}{\bibinfo{person}{Thong Hoang}, \bibinfo{person}{Hoa~Khanh
  Dam}, \bibinfo{person}{Yasutaka Kamei}, \bibinfo{person}{David Lo}, {and}
  \bibinfo{person}{Naoyasu Ubayashi}.} \bibinfo{year}{2019}\natexlab{}.
\newblock \showarticletitle{{DeepJIT}: An End-to-End Deep Learning Framework
  for Just-in-Time Defect Prediction}. In \bibinfo{booktitle}{\emph{2019
  {IEEE}/{ACM} 16th International Conference on Mining Software Repositories
  ({MSR})}}. \bibinfo{publisher}{{IEEE}}.
\newblock
\urldef\tempurl%
\url{https://doi.org/10.1109/msr.2019.00016}
\showDOI{\tempurl}


\bibitem[Hoang et~al\mbox{.}(2020)]%
        {Hoang2020}
\bibfield{author}{\bibinfo{person}{Thong Hoang}, \bibinfo{person}{Hong~Jin
  Kang}, \bibinfo{person}{David Lo}, {and} \bibinfo{person}{Julia Lawall}.}
  \bibinfo{year}{2020}\natexlab{}.
\newblock \showarticletitle{{CC}2Vec}. In \bibinfo{booktitle}{\emph{Proceedings
  of the {ACM}/{IEEE} 42nd International Conference on Software Engineering}}.
  \bibinfo{publisher}{{ACM}}.
\newblock
\urldef\tempurl%
\url{https://doi.org/10.1145/3377811.3380361}
\showDOI{\tempurl}


\bibitem[Huang et~al\mbox{.}(2017)]%
        {Huang2017}
\bibfield{author}{\bibinfo{person}{Qiao Huang}, \bibinfo{person}{Xin Xia},
  {and} \bibinfo{person}{David Lo}.} \bibinfo{year}{2017}\natexlab{}.
\newblock \showarticletitle{Supervised vs Unsupervised Models: A Holistic Look
  at Effort-Aware Just-in-Time Defect Prediction}. In
  \bibinfo{booktitle}{\emph{2017 {IEEE} International Conference on Software
  Maintenance and Evolution ({ICSME})}}. \bibinfo{publisher}{{IEEE}}.
\newblock


\bibitem[Huang et~al\mbox{.}(2018)]%
        {Huang2019c}
\bibfield{author}{\bibinfo{person}{Qiao Huang}, \bibinfo{person}{Xin Xia},
  {and} \bibinfo{person}{David Lo}.} \bibinfo{year}{2018}\natexlab{}.
\newblock \showarticletitle{Revisiting supervised and unsupervised models for
  effort-aware just-in-time defect prediction}.
\newblock \bibinfo{journal}{\emph{Empirical Software Engineering}}
  \bibinfo{volume}{24}, \bibinfo{number}{5} (\bibinfo{date}{Oct.}
  \bibinfo{year}{2018}), \bibinfo{pages}{2823--2862}.
\newblock
\urldef\tempurl%
\url{https://doi.org/10.1007/s10664-018-9661-2}
\showDOI{\tempurl}


\bibitem[Kamei et~al\mbox{.}(2013)]%
        {Kamei2013a}
\bibfield{author}{\bibinfo{person}{Yasutaka Kamei}, \bibinfo{person}{Emad
  Shihab}, \bibinfo{person}{Bram Adams}, \bibinfo{person}{Ahmed~E. Hassan},
  \bibinfo{person}{Audris Mockus}, \bibinfo{person}{Anand Sinha}, {and}
  \bibinfo{person}{Naoyasu Ubayashi}.} \bibinfo{year}{2013}\natexlab{}.
\newblock \showarticletitle{A large-scale empirical study of just-in-time
  quality assurance}.
\newblock \bibinfo{journal}{\emph{{IEEE} Transactions on Software Engineering}}
  \bibinfo{volume}{39}, \bibinfo{number}{6} (\bibinfo{date}{June}
  \bibinfo{year}{2013}), \bibinfo{pages}{757--773}.
\newblock
\urldef\tempurl%
\url{https://doi.org/10.1109/tse.2012.70}
\showDOI{\tempurl}


\bibitem[Kim et~al\mbox{.}(2008)]%
        {Kim2008}
\bibfield{author}{\bibinfo{person}{Sunghun Kim}, \bibinfo{person}{E.~James
  Whitehead}, {and} \bibinfo{person}{Yi Zhang}.}
  \bibinfo{year}{2008}\natexlab{}.
\newblock \showarticletitle{Classifying Software Changes: Clean or Buggy?}
\newblock \bibinfo{journal}{\emph{{IEEE} Transactions on Software Engineering}}
  \bibinfo{volume}{34}, \bibinfo{number}{2} (\bibinfo{date}{March}
  \bibinfo{year}{2008}), \bibinfo{pages}{181--196}.
\newblock
\urldef\tempurl%
\url{https://doi.org/10.1109/tse.2007.70773}
\showDOI{\tempurl}


\bibitem[Kononenko et~al\mbox{.}(2015)]%
        {Kononenko2015}
\bibfield{author}{\bibinfo{person}{Oleksii Kononenko}, \bibinfo{person}{Olga
  Baysal}, \bibinfo{person}{Latifa Guerrouj}, \bibinfo{person}{Yaxin Cao},
  {and} \bibinfo{person}{Michael~W. Godfrey}.} \bibinfo{year}{2015}\natexlab{}.
\newblock \showarticletitle{Investigating code review quality: Do people and
  participation matter?}. In \bibinfo{booktitle}{\emph{2015 {IEEE}
  International Conference on Software Maintenance and Evolution ({ICSME})}}.
  \bibinfo{publisher}{{IEEE}}.
\newblock
\urldef\tempurl%
\url{https://doi.org/10.1109/icsm.2015.7332457}
\showDOI{\tempurl}


\bibitem[Liu et~al\mbox{.}(2017)]%
        {Liu2017}
\bibfield{author}{\bibinfo{person}{Jinping Liu}, \bibinfo{person}{Yuming Zhou},
  \bibinfo{person}{Yibiao Yang}, \bibinfo{person}{Hongmin Lu}, {and}
  \bibinfo{person}{Baowen Xu}.} \bibinfo{year}{2017}\natexlab{}.
\newblock \showarticletitle{Code Churn: A Neglected Metric in Effort-Aware
  Just-in-Time Defect Prediction}. In \bibinfo{booktitle}{\emph{2017
  {ACM}/{IEEE} International Symposium on Empirical Software Engineering and
  Measurement ({ESEM})}}. \bibinfo{publisher}{{IEEE}}.
\newblock
\urldef\tempurl%
\url{https://doi.org/10.1109/esem.2017.8}
\showDOI{\tempurl}


\bibitem[Majumder et~al\mbox{.}(2018)]%
        {Majumder2018}
\bibfield{author}{\bibinfo{person}{Suvodeep Majumder}, \bibinfo{person}{Nikhila
  Balaji}, \bibinfo{person}{Katie Brey}, \bibinfo{person}{Wei Fu}, {and}
  \bibinfo{person}{Tim Menzies}.} \bibinfo{year}{2018}\natexlab{}.
\newblock \showarticletitle{500+ times faster than deep learning}. In
  \bibinfo{booktitle}{\emph{Proceedings of the 15th International Conference on
  Mining Software Repositories}}. \bibinfo{publisher}{{ACM}}.
\newblock


\bibitem[Martin(2009)]%
        {martin2009clean}
\bibfield{author}{\bibinfo{person}{Robert~C Martin}.}
  \bibinfo{year}{2009}\natexlab{}.
\newblock \bibinfo{booktitle}{\emph{Clean code: a handbook of agile software
  craftsmanship}}.
\newblock \bibinfo{publisher}{Pearson Education}. 53--74 pages.
\newblock


\bibitem[McCabe(1976)]%
        {1702388}
\bibfield{author}{\bibinfo{person}{T.J. McCabe}.}
  \bibinfo{year}{1976}\natexlab{}.
\newblock \showarticletitle{A Complexity Measure}.
\newblock \bibinfo{journal}{\emph{IEEE Transactions on Software Engineering}}
  \bibinfo{volume}{SE-2}, \bibinfo{number}{4} (\bibinfo{year}{1976}),
  \bibinfo{pages}{308--320}.
\newblock
\urldef\tempurl%
\url{https://doi.org/10.1109/TSE.1976.233837}
\showDOI{\tempurl}


\bibitem[McIntosh and Kamei(2018)]%
        {McIntosh2018a}
\bibfield{author}{\bibinfo{person}{Shane McIntosh} {and}
  \bibinfo{person}{Yasutaka Kamei}.} \bibinfo{year}{2018}\natexlab{}.
\newblock \showarticletitle{Are Fix-Inducing Changes a Moving Target? A
  Longitudinal Case Study of Just-In-Time Defect Prediction}.
\newblock \bibinfo{journal}{\emph{{IEEE} Transactions on Software Engineering}}
  \bibinfo{volume}{44}, \bibinfo{number}{5} (\bibinfo{date}{May}
  \bibinfo{year}{2018}), \bibinfo{pages}{412--428}.
\newblock


\bibitem[Mockus and Weiss(2002)]%
        {Mockus2000}
\bibfield{author}{\bibinfo{person}{Audris Mockus} {and}
  \bibinfo{person}{David~M. Weiss}.} \bibinfo{year}{2002}\natexlab{}.
\newblock \showarticletitle{Predicting risk of software changes}.
\newblock \bibinfo{journal}{\emph{Bell Labs Technical Journal}}
  \bibinfo{volume}{5}, \bibinfo{number}{2} (\bibinfo{date}{Aug.}
  \bibinfo{year}{2002}), \bibinfo{pages}{169--180}.
\newblock
\urldef\tempurl%
\url{https://doi.org/10.1002/bltj.2229}
\showDOI{\tempurl}


\bibitem[Moser et~al\mbox{.}(2008)]%
        {Moser2008a}
\bibfield{author}{\bibinfo{person}{Raimund Moser}, \bibinfo{person}{Witold
  Pedrycz}, {and} \bibinfo{person}{Giancarlo Succi}.}
  \bibinfo{year}{2008}\natexlab{}.
\newblock \showarticletitle{A comparative analysis of the efficiency of change
  metrics and static code attributes for defect prediction}. In
  \bibinfo{booktitle}{\emph{Proceedings of the 13th international conference on
  Software engineering - {ICSE} {\textquotesingle}08}}.
  \bibinfo{publisher}{{ACM} Press}.
\newblock
\urldef\tempurl%
\url{https://doi.org/10.1145/1368088.1368114}
\showDOI{\tempurl}


\bibitem[Nayrolles and Hamou-Lhadj(2018)]%
        {Nayrolles2018}
\bibfield{author}{\bibinfo{person}{Mathieu Nayrolles} {and}
  \bibinfo{person}{Abdelwahab Hamou-Lhadj}.} \bibinfo{year}{2018}\natexlab{}.
\newblock \showarticletitle{{CLEVER: combining code metrics with clone
  detection for just-in-time fault prevention and resolution in large
  industrial projects}}. In \bibinfo{booktitle}{\emph{Proceedings of the 15th
  International Conference on Mining Software Repositories}}.
  \bibinfo{publisher}{{ACM}}.
\newblock
\urldef\tempurl%
\url{https://doi.org/10.1145/3196398.3196438}
\showDOI{\tempurl}


\bibitem[Osman et~al\mbox{.}(2017)]%
        {Osman2017}
\bibfield{author}{\bibinfo{person}{Haidar Osman}, \bibinfo{person}{Mohammad
  Ghafari}, \bibinfo{person}{Oscar Nierstrasz}, {and} \bibinfo{person}{Mircea
  Lungu}.} \bibinfo{year}{2017}\natexlab{}.
\newblock \showarticletitle{An Extensive Analysis of Efficient Bug Prediction
  Configurations}.
\newblock \bibinfo{journal}{\emph{Proceedings of the 13th International
  Conference on Predictive Models and Data Analytics in Software Engineering -
  PROMISE}} (\bibinfo{year}{2017}).
\newblock


\bibitem[Ostrand et~al\mbox{.}(2005)]%
        {Ostrand2005}
\bibfield{author}{\bibinfo{person}{T.J. Ostrand}, \bibinfo{person}{E.J.
  Weyuker}, {and} \bibinfo{person}{R.M. Bell}.}
  \bibinfo{year}{2005}\natexlab{}.
\newblock \showarticletitle{Predicting the location and number of faults in
  large software systems}.
\newblock \bibinfo{journal}{\emph{{IEEE} Transactions on Software Engineering}}
  \bibinfo{volume}{31}, \bibinfo{number}{4} (\bibinfo{date}{April}
  \bibinfo{year}{2005}), \bibinfo{pages}{340--355}.
\newblock
\urldef\tempurl%
\url{https://doi.org/10.1109/tse.2005.49}
\showDOI{\tempurl}


\bibitem[Pornprasit and Tantithamthavorn(2021)]%
        {Pornprasit2021ARXIV}
\bibfield{author}{\bibinfo{person}{Chanathip Pornprasit} {and}
  \bibinfo{person}{Chakkrit Tantithamthavorn}.}
  \bibinfo{year}{2021}\natexlab{}.
\newblock \showarticletitle{JITLine: A Simpler, Better, Faster, Finer-grained
  Just-In-Time Defect Prediction}.
\newblock  (\bibinfo{year}{2021}).
\newblock
\urldef\tempurl%
\url{https://doi.org/10.48550/arXiv.2103.07068}
\showURL{%
\tempurl}


\bibitem[Scott and Knott(1974)]%
        {10.2307/2529204}
\bibfield{author}{\bibinfo{person}{A.~J. Scott} {and} \bibinfo{person}{M.
  Knott}.} \bibinfo{year}{1974}\natexlab{}.
\newblock \showarticletitle{A Cluster Analysis Method for Grouping Means in the
  Analysis of Variance}.
\newblock \bibinfo{journal}{\emph{Biometrics}} \bibinfo{volume}{30},
  \bibinfo{number}{3} (\bibinfo{year}{1974}), \bibinfo{pages}{507--512}.
\newblock
\showISSN{0006341X, 15410420}


\bibitem[Shihab et~al\mbox{.}(2012)]%
        {Shihab2012}
\bibfield{author}{\bibinfo{person}{Emad Shihab}, \bibinfo{person}{Ahmed~E.
  Hassan}, \bibinfo{person}{Bram Adams}, {and} \bibinfo{person}{Zhen~Ming
  Jiang}.} \bibinfo{year}{2012}\natexlab{}.
\newblock \showarticletitle{An industrial study on the risk of software
  changes}. In \bibinfo{booktitle}{\emph{Proceedings of the {ACM} {SIGSOFT}
  20th International Symposium on the Foundations of Software Engineering -
  {FSE} {\textquotesingle}12}}. \bibinfo{publisher}{{ACM} Press}.
\newblock
\urldef\tempurl%
\url{https://doi.org/10.1145/2393596.2393670}
\showDOI{\tempurl}


\bibitem[{\'{S}}liwerski et~al\mbox{.}(2005)]%
        {Jaceksliwerski2005}
\bibfield{author}{\bibinfo{person}{Jacek {\'{S}}liwerski},
  \bibinfo{person}{Thomas Zimmermann}, {and} \bibinfo{person}{Andreas Zeller}.}
  \bibinfo{year}{2005}\natexlab{}.
\newblock \showarticletitle{When do changes induce fixes?}
\newblock \bibinfo{journal}{\emph{{ACM} {SIGSOFT} Software Engineering Notes}}
  \bibinfo{volume}{30}, \bibinfo{number}{4} (\bibinfo{date}{July}
  \bibinfo{year}{2005}), \bibinfo{pages}{1--5}.
\newblock
\urldef\tempurl%
\url{https://doi.org/10.1145/1082983.1083147}
\showDOI{\tempurl}


\bibitem[Sola and Sevilla(1997)]%
        {589532}
\bibfield{author}{\bibinfo{person}{J. Sola} {and} \bibinfo{person}{J.
  Sevilla}.} \bibinfo{year}{1997}\natexlab{}.
\newblock \showarticletitle{Importance of input data normalization for the
  application of neural networks to complex industrial problems}.
\newblock \bibinfo{journal}{\emph{IEEE Transactions on Nuclear Science}}
  \bibinfo{volume}{44}, \bibinfo{number}{3} (\bibinfo{year}{1997}),
  \bibinfo{pages}{1464--1468}.
\newblock
\urldef\tempurl%
\url{https://doi.org/10.1109/23.589532}
\showDOI{\tempurl}


\bibitem[Tantithamthavorn et~al\mbox{.}(2017)]%
        {scottknott}
\bibfield{author}{\bibinfo{person}{Chakkrit Tantithamthavorn},
  \bibinfo{person}{Shane McIntosh}, \bibinfo{person}{Ahmed~E. Hassan}, {and}
  \bibinfo{person}{Kenichi Matsumoto}.} \bibinfo{year}{2017}\natexlab{}.
\newblock \showarticletitle{An Empirical Comparison of Model Validation
  Techniques for Defect Prediction Models}.
\newblock  \bibinfo{number}{1} (\bibinfo{year}{2017}).
\newblock


\bibitem[Wilcoxon(1945)]%
        {10.2307/3001968}
\bibfield{author}{\bibinfo{person}{Frank Wilcoxon}.}
  \bibinfo{year}{1945}\natexlab{}.
\newblock \showarticletitle{Individual Comparisons by Ranking Methods}.
\newblock \bibinfo{journal}{\emph{Biometrics Bulletin}} \bibinfo{volume}{1},
  \bibinfo{number}{6} (\bibinfo{year}{1945}), \bibinfo{pages}{80--83}.
\newblock
\showISSN{00994987}


\bibitem[Yadav et~al\mbox{.}(2018)]%
        {Yadav2018}
\bibfield{author}{\bibinfo{person}{Monika Yadav}, \bibinfo{person}{Vijendra
  Singh}, {and} \bibinfo{person}{Priyanka Rastogi}.}
  \bibinfo{year}{2018}\natexlab{}.
\newblock \showarticletitle{Deep Learning for Software Defect Prediction in
  time}. In \bibinfo{booktitle}{\emph{2018 Fifth International Conference on
  Parallel, Distributed and Grid Computing ({PDGC})}}.
  \bibinfo{publisher}{{IEEE}}.
\newblock


\bibitem[Yan et~al\mbox{.}(2020)]%
        {Yan2020}
\bibfield{author}{\bibinfo{person}{Meng Yan}, \bibinfo{person}{Xin Xia},
  \bibinfo{person}{Yuanrui Fan}, \bibinfo{person}{David Lo},
  \bibinfo{person}{Ahmed~E. Hassan}, {and} \bibinfo{person}{Xindong Zhang}.}
  \bibinfo{year}{2020}\natexlab{}.
\newblock \showarticletitle{Effort-aware just-in-time defect identification in
  practice: a case study at Alibaba}. In \bibinfo{booktitle}{\emph{Proceedings
  of the 28th {ACM} Joint Meeting on European Software Engineering Conference
  and Symposium on the Foundations of Software Engineering}}.
  \bibinfo{publisher}{{ACM}}.
\newblock
\urldef\tempurl%
\url{https://doi.org/10.1145/3368089.3417048}
\showDOI{\tempurl}


\bibitem[Yan et~al\mbox{.}(2019)]%
        {Yan2019}
\bibfield{author}{\bibinfo{person}{Meng Yan}, \bibinfo{person}{Xin Xia},
  \bibinfo{person}{David Lo}, \bibinfo{person}{Ahmed~E. Hassan}, {and}
  \bibinfo{person}{Shanping Li}.} \bibinfo{year}{2019}\natexlab{}.
\newblock \showarticletitle{Characterizing and identifying reverted commits}.
\newblock \bibinfo{journal}{\emph{Empirical Software Engineering}}
  \bibinfo{volume}{24}, \bibinfo{number}{4} (\bibinfo{date}{March}
  \bibinfo{year}{2019}), \bibinfo{pages}{2171--2208}.
\newblock
\urldef\tempurl%
\url{https://doi.org/10.1007/s10664-019-09688-8}
\showDOI{\tempurl}


\bibitem[Yang et~al\mbox{.}(2017)]%
        {Yang2017}
\bibfield{author}{\bibinfo{person}{Xinli Yang}, \bibinfo{person}{David Lo},
  \bibinfo{person}{Xin Xia}, {and} \bibinfo{person}{Jianling Sun}.}
  \bibinfo{year}{2017}\natexlab{}.
\newblock \showarticletitle{{TLEL}: A two-layer ensemble learning approach for
  just-in-time defect prediction}.
\newblock \bibinfo{journal}{\emph{Information and Software Technology}}
  \bibinfo{volume}{87} (\bibinfo{date}{July} \bibinfo{year}{2017}),
  \bibinfo{pages}{206--220}.
\newblock
\urldef\tempurl%
\url{https://doi.org/10.1016/j.infsof.2017.03.007}
\showDOI{\tempurl}


\bibitem[Yang et~al\mbox{.}(2015)]%
        {Yang2015}
\bibfield{author}{\bibinfo{person}{Xinli Yang}, \bibinfo{person}{David Lo},
  \bibinfo{person}{Xin Xia}, \bibinfo{person}{Yun Zhang}, {and}
  \bibinfo{person}{Jianling Sun}.} \bibinfo{year}{2015}\natexlab{}.
\newblock \showarticletitle{Deep Learning for Just-in-Time Defect Prediction}.
  In \bibinfo{booktitle}{\emph{2015 {IEEE} International Conference on Software
  Quality, Reliability and Security}}. \bibinfo{publisher}{{IEEE}}.
\newblock


\bibitem[Yang et~al\mbox{.}(2016)]%
        {Yang2016}
\bibfield{author}{\bibinfo{person}{Yibiao Yang}, \bibinfo{person}{Yuming Zhou},
  \bibinfo{person}{Jinping Liu}, \bibinfo{person}{Yangyang Zhao},
  \bibinfo{person}{Hongmin Lu}, \bibinfo{person}{Lei Xu},
  \bibinfo{person}{Baowen Xu}, {and} \bibinfo{person}{Hareton Leung}.}
  \bibinfo{year}{2016}\natexlab{}.
\newblock \showarticletitle{Effort-aware just-in-time defect prediction: simple
  unsupervised models could be better than supervised models}. In
  \bibinfo{booktitle}{\emph{Proceedings of the 2016 24th {ACM} {SIGSOFT}
  International Symposium on Foundations of Software Engineering}}.
  \bibinfo{publisher}{{ACM}}.
\newblock
\urldef\tempurl%
\url{https://doi.org/10.1145/2950290.2950353}
\showDOI{\tempurl}


\bibitem[Yao and Shepperd(2020)]%
        {10.1145/3383219.3383232}
\bibfield{author}{\bibinfo{person}{Jingxiu Yao} {and} \bibinfo{person}{Martin
  Shepperd}.} \bibinfo{year}{2020}\natexlab{}.
\newblock \showarticletitle{Assessing Software Defection Prediction
  Performance: Why Using the Matthews Correlation Coefficient Matters}. In
  \bibinfo{booktitle}{\emph{Proceedings of the Evaluation and Assessment in
  Software Engineering}} (Trondheim, Norway) \emph{(\bibinfo{series}{EASE
  '20})}. \bibinfo{publisher}{Association for Computing Machinery},
  \bibinfo{address}{New York, NY, USA}, \bibinfo{pages}{120–129}.
\newblock
\showISBNx{9781450377317}
\urldef\tempurl%
\url{https://doi.org/10.1145/3383219.3383232}
\showDOI{\tempurl}


\bibitem[Young et~al\mbox{.}(2018)]%
        {Young2018}
\bibfield{author}{\bibinfo{person}{Steven Young}, \bibinfo{person}{Tamer
  Abdou}, {and} \bibinfo{person}{Ayse Bener}.} \bibinfo{year}{2018}\natexlab{}.
\newblock \showarticletitle{A Replication Study: Just-in-Time Defect Prediction
  with Ensemble Learning}. In \bibinfo{booktitle}{\emph{Proceedings of the 6th
  International Workshop on Realizing Artificial Intelligence Synergies in
  Software Engineering}} (Gothenburg, Sweden) \emph{(\bibinfo{series}{RAISE
  '18})}. \bibinfo{publisher}{Association for Computing Machinery},
  \bibinfo{address}{New York, NY, USA}, \bibinfo{pages}{42–47}.
\newblock
\showISBNx{9781450357234}
\urldef\tempurl%
\url{https://dl.acm.org/doi/10.1145/3194104.3194110}
\showURL{%
\tempurl}


\bibitem[Zeng et~al\mbox{.}(2021)]%
        {Zeng2021}
\bibfield{author}{\bibinfo{person}{Zhengran Zeng}, \bibinfo{person}{Yuqun
  Zhang}, \bibinfo{person}{Haotian Zhang}, {and} \bibinfo{person}{Lingming
  Zhang}.} \bibinfo{year}{2021}\natexlab{}.
\newblock \showarticletitle{Deep just-in-time defect prediction: how far are
  we?}. In \bibinfo{booktitle}{\emph{Proceedings of the 30th {ACM} {SIGSOFT}
  International Symposium on Software Testing and Analysis}}.
  \bibinfo{publisher}{{ACM}}.
\newblock


\end{thebibliography}

\end{document}